\documentclass{article}

\usepackage{amssymb}
\usepackage{graphicx}% Include figure files
\usepackage{dcolumn}% Align table columns on decimal point
\usepackage{bm}% bold math
\date{}
\begin{document}

\title{\bf Nonsingular universe from generalized
thermostatistics}
\author{\\{B. Vakili}$^{1,}$\thanks{email: b.vakili@iauctb.ac.ir } \nonumber\
and \nonumber\ {M. A. Gorji}$^{2,}$\thanks{email: m.gorji@stu.umz.ac.ir}\\\\
{\small $^1${\it Department of Physics, Central Tehran Branch,
Islamic Azad University, Tehran, Iran}}
\\{\small $^2${\it Department of Physics, University of Mazandaran, P.O.
Box 47416-95447, Babolsar, Iran}}}

\maketitle
\begin{abstract}We study the statistical mechanics of the early radiation
dominated universe in the context of a generalized uncertainty principle
which supports the existence of a minimal length scale. Utilizing the
resultant modified thermodynamical quantities, we obtain a deformed
Friedmann equation which is very similar to that arises from loop quantum
cosmology scenarios. The energy and entropy densities get maximum bounds
about Planck temperature and a nonsingular universe then emerges in this
setup.\\
\begin{description}
\item[PACS numbers]
04.60.Bc, 98.80.Cq
\item[Key Words]
Quantum Gravity Phenomenology, Thermodynamics, Cosmology
\end{description}
\end{abstract}
\section{Introduction}
Although the standard Big Bang model of cosmology describes our
observable universe with high accuracy, its shortages are revealed
when, for instance, the model is implemented to describe the early
state of the universe. The initial condition (fine-tuning) problem
emerges in the ultraviolet regime which can be resolved by
considering an accelerated inflationary phase about $10^{15}$ GeV
\cite{Inflation}. Furthermore, the universe starts from an initial
Big Bang singularity where the classical equation of general
relativity fails to be applicable. The Big Bang singularity,
however, appears at very high energy regime where the quantum
effects will become significant and it is then natural to expect
that the problem will be resolved in the framework of a quantum
theory of gravity. Unlike electromagnetic, weak and strong nuclear
interactions which are quantized very successfully by means of the
gauge field quantization method, quantization of gravity is not an
easy task. While the three other forces are classically described by
the gauge fields that propagate on the fixed Minkowski background
geometry, gravity itself is a background geometry and quantization
of gravity is then equivalent to the quantization of geometry.
Although a full quantum theory of gravity is not formulated yet, the
elegant work of Ashtekar in Ref. \cite{Ashtekar} inspired the loop
representation of the quantum general relativity \cite{LQG} (for
review see Ref. \cite{RLQG}). Very interestingly, the Big Bang
singularities are replaced by quantum bounces when loop quantum
gravity method is applied into the homogeneous and isotropic
space-time in loop quantum cosmology scenario \cite{LQC} (see Ref.
\cite{RLQC} for review). Indeed, the volume operator is quantized
with respect to a fundamental length scale (preferably of the order
of Planck length) in loop quantum gravity \cite{LQG-VO} and, then,
an ultraviolet cutoff appears which removes the Big Bang
singularity. In fact, existence of a minimal length scale is a
common address of any quantum theory of gravity such as string
theory and loop quantum gravity \cite{String}. For instance,
inspired by the string theory, the generalized uncertainty principle
(GUP) is investigated which supports the existence of a minimal
length through a nonzero uncertainty in position measurement
\cite{GUP,GUPUV,GUPUV2}. The statistical mechanics then will be
significantly modified in the high energy regimes and the
thermodynamics of the early universe will be affected by the GUP
corrections \cite{GUP}-\cite{GUP-THR2}. Since the GUP effects are
related to the existence of a minimal length cutoff, it is natural
to expect that this setup may resolve the Big Bang singularity
\cite{GUP-NS,Ali,Ali2}. Motivated by this idea, we study the effects
of GUP on the thermodynamics of the early universe and interestingly
we find a modified Friedmann equation which is very similar to that
arises from loop quantum cosmology.

The structure of the paper is as follows: In section 2, we formulate
the generalized uncertainty principle and the idea of the existence
of a minimal length from which we find the associated deformed
density of states. In section 3, we will study the thermodynamics of
the early radiation dominated universe in GUP framework by means of
the deformed density of states. In section 4, the modified Friedmann
equations are obtained and the results are compared with those arise
from loop quantum cosmology scenarios. Section 5 is devoted to the
summary and conclusions.

\section{GUP and deformed density of states}
Since the standard Heisenberg uncertainty principle does not take
into account any universal length scale, considering a minimal
length scale will naturally lead to the modification of this
principle. Nevertheless, there is not a unique deformation to the
uncertainty principle. Here, we consider the generalized uncertainty
relation as\footnote{We work in natural units $\hbar=c=k_{_B}=1$,
where $\hbar$, $c$ and $k_{_B}$ are the Planck constant, speed of
light in vacuum and Boltzman constant, respectively.}
\begin{equation}\label{GUP0}
\Delta q\Delta p\geq\,\frac{1}{2}\big(1+(\beta\Delta p)^{2n}\big)\,,
\end{equation}
where  $n\geq1$, $\beta={\mathcal O}(1)\,l_{_{\rm Pl}}$, with
$l_{_{\rm Pl}}$ being the Planck length, is the deformation
parameter and the dimensionless numerical coefficient ${\mathcal
O}(1)$ should be fixed by the experiments \cite{experiment}. In the
$n=1$ case, relation (\ref{GUP0}) reduces to the well-known
generalized uncertainty relation which is investigated in the
context of the string theory \cite{GUPUV}. Apart from some details
of the various versions of such uncertainty relations, all of them
should satisfy at least two properties: They should support the idea
of existence of a minimal length scale and the corresponding
uncertainty relations have to reduce to the standard Heisenberg
uncertainty principle in the light of the correspondence principle.
By a little algebra, it is easy to show that the proposed model
(\ref{GUP0}) predicts a minimal uncertainty in position measurement
as
\begin{equation}\label{GUP-min}
\Delta q_{\rm min}=2n(2n-1)^{\frac{1-2n}{2n}}\,\beta,
\end{equation}
from which it is seen that the uncertainty relation (\ref{GUP0})
results to a cutoff on minimal length. Also, it clearly reduces to
the standard uncertainty principle in the low energy limit
$\beta\rightarrow\,0$.

By using of the standard (quantization scheme independent)
definition $(\Delta A)^2\, (\Delta
B)^2\geq\frac{1}{4}\langle[A,B]\rangle^2$, it is straightforward to
show that the modified uncertainty relation (\ref{GUP0}) can be
realized from the following deformed Heisenberg algebra
\begin{eqnarray}\label{GUP}
[\hat{q},\hat{p}]=(1+(\beta{\hat p})^{2n}).
\end{eqnarray}
However, since in this work we are interested in the statistical
mechanics of a many-particle system, we generalize this setup to the
three dimensional case. A natural generalization which preserve the
rotational symmetry will be
\begin{eqnarray}\label{GUP1}
[\hat{q}_i,\hat{p}_j]=(1+(\beta{\hat p})^{2n})\,\delta_{ij},
\end{eqnarray}
where ${\hat p}^2={\hat p}_{_x}^2+{\hat p}_{_y}^2+{\hat p}_{_z}^2$.
Because of the existence of a nonzero uncertainty in positions
measurement, it is natural to suppose
\begin{eqnarray}\label{GUP2}
[\hat{p}_i,\hat{p}_j]=0,
\end{eqnarray}
which allows us to work in momentum representation with
\begin{eqnarray}\label{GUP-HS}
\hat{p}_i.\psi(p)=p_i\psi(p),\hspace{.7cm}\hat{q}_i.\psi(p)
=i(1+(\beta p)^{2n})\frac{\partial\psi(p)}{\partial p_i},
\end{eqnarray}
where $p=\sqrt{p_{_x}^2+p_{_y}^2+p_{_z}^2}$. The commutation relation between the
position operators for the above representation then becomes
\begin{eqnarray}\label{GUP3}
[\hat{q}_i,\hat{q}_j]=2in\beta^{2n}\hat{p}^{2n-2}(
\hat{p}_i\hat{q}_j-\hat{p}_j\hat{q}_i),
\end{eqnarray}
where the results of the Ref. \cite{GUPUV} can be correctly
reproduced for the case $n=1$.

In order to obtain the density of states of a statistical system, we
consider the classical limit of this setup. The deformed Heisenberg
algebra which is defined by the commutation relations (\ref{GUP1}),
(\ref{GUP2}) and (\ref{GUP3}) then can be extracted from the
deformed non-canonical Poisson algebra
\begin{eqnarray}\label{DPA}
\{q_i,q_j\}=2n\beta^{2n}p^{2n-2}\big(p_iq_j-p_jq_i\big),\hspace{1cm}
{\{q_i,p_j\}=\big(1+(\beta p)^{2n}\big)\,\delta_{ij}},\hspace{1cm}\{p_i,p_j\}=0,
\end{eqnarray}
in which the operators are replaced by their corresponding phase
space variables counterparts ($q_i$ and $p_i$) and the Dirac
commutators by the Poisson brackets. The invariant Liouville measure
which is consistent with the non-canonical Poisson algebra
(\ref{DPA}) then will be \cite{GUP-DOS,GUP-Fityo,GUP-THR2}
\begin{equation}\label{DPSV}
d^3q\,d^3p\rightarrow\,\frac{d^3q\,d^3p}{\big(1+(\beta{p})^{2n}\big)^3}\,.
\end{equation}
In Ref. \cite{DOS-LT}, we have obtained the deformed Liouville
measure for a generally deformed Poisson algebra and showed that it
is invariant under the time evolution of the system which guaranties
the validity of the Liouville theorem in such theories.

\section{Thermodynamics of the early universe}
Having the deformed invariant Liouville measure (\ref{DPSV}) at
hand, one can easily study the statistical mechanics in
semiclassical regime in the context of the GUP (\ref{GUP0}).

By assuming a radiation dominated universe in early times of cosmic
evolution, we consider the thermodynamics of the ultra-relativistic
gas confined in volume $V$ at temperature $T$. The pressure is given
by the standard definition
\begin{equation}\label{Pdef}
{\mathrm{P}}V=\mp\,T\sum_{\varepsilon}\ln\big(1\mp\,e^{-\varepsilon/T}\big)\,,
\end{equation}
where the signs $(-)$ and $(+)$ refer to the bosons and fermions
respectively and the summation is over the quantized energies of the
microstates $\varepsilon$. At high temperature regime, the summation
over $\varepsilon$ may be replaced by an integral over all phase
space variables in the semiclassical regime by means of the
invariant Liouville measure (\ref{DPSV}) as
\begin{equation}\label{DDS}
\sum_{\varepsilon}\rightarrow\,\frac{g\,V}{(2\pi)^3}
\int\frac{d^3{p}}{\big(1+(\beta{p})^{2n}\big)^3}\,,
\end{equation}
where $g$ is the number of relativistic degrees of freedom and we
have also performed the integral over the positions as
$\int_{V}d^3{q}=V$. Now, the pressure in the GUP framework can be
obtained from definition (\ref{Pdef}) and the deformed density of
states (\ref{DDS}) as
\begin{eqnarray}\label{QGP1}
{\mathrm{P}}_{\mp}=\mp\frac{g_{_{\mp}}T}{2\pi^2}\int_{0}^{\infty}
\frac{\ln\big(1{\mp}e^{-p/T}\big)\,p^2dp}{\big(1+(\beta{p})^{2n}\big)^3}\,,
\end{eqnarray}
where $g_{_{-}}$ and $g_{_{+}}$ are the number of relativistic degrees of
freedom for bosons and fermions respectively and we have substituted
$\varepsilon=p$ for the ultra-relativistic particles. Using the
dimensionless variable $x=p/T$, the relation (\ref{QGP1}) becomes
\begin{eqnarray}\label{P-def}
{\mathrm{P}}_{\mp}(T)=\frac{{\pi^2}T^4}{90}\,I_{\mp}(T)\,,
\end{eqnarray}
in which we have defined
\begin{eqnarray}\label{I2}
I_{\mp}(T)=\mp\frac{45}{\pi^4}\,g_{_{\mp}}\int_{0}^{\infty}
\frac{\ln\big(1\mp\,e^{-x}\big)\,x^2dx}{\big(1+(\beta{xT})^{
2n}\big)^3}\,.
\end{eqnarray}
To evaluate the above integrals we may expand the denominator for
$\beta{p}=\beta{xT}\ll1$, for which by using of the integral formula
\[\int_{0}^{\infty}\frac{x^{s-1}\,dx}{e^x+1}
=\big(1-\frac{1}{2^{s-1}}\big)\,\int_{0}^{\infty}\frac{x^{s-1}\,
dx}{e^x-1},\] the integrals in (\ref{I2}) can be evaluated as
\begin{equation}\label{I-}
I_{-}(T)=g_{_{-}}\left[1-\frac{135}{\pi^4}\Gamma(2n+3)\zeta(2n+4)
\beta^{2n}\,T^{2n}\right]\,,
\end{equation}
\begin{equation}\label{I+}
I_{+}(T)=\frac{7}{8}g_{_{+}}\left[1-\frac{1080}{7\pi^4}\Gamma(2n+3)
\zeta(2n+4)\Big(1-2^{-(2n+3)}\Big)\beta^{2n}\,T^{2n}\right]\,.
\end{equation}
These functions satisfy the conditions
$I_{-}(T\rightarrow\,0)=g_{_{-}}$, and $I_{+
}(T\rightarrow\,0)=\frac{7}{8}g_{_{+}}$. Thus, the usual results of
the low temperature regime are recovered. Upon Substitution of
(\ref{I-}) and (\ref{I+}) into relation (\ref{P-def}), we get the
total pressure ${\mathrm{P}}(T)={\mathrm{P}}_{
-}(T)+{\mathrm{P}}_{+}(T)$ as
\begin{equation}\label{P}
{\mathrm{P}}(T)={\mathrm{P}}_0(T)\left[1-\chi(n)\frac{g_{\ast}(n)}{g_{0\ast}
}\beta_0^{2n}\bigg(\frac{T}{T_{\rm Pl}}\bigg)^{2n}\right]\,,
\end{equation}
where ${\mathrm{P}}_0=\frac{\pi^2g_{0\ast}}{90}\,T^4$ is the standard
pressure for the ultra-relativistic gas, $g_{0\ast}=g_{_{-}}+\frac{7
}{8}g_{_{+}}$ is the number of effective degrees of freedom, and we
have also defined
\begin{equation}\label{chi}
\chi(n)=\frac{135}{\pi^4}\Gamma(2n+3)\zeta(2n+4)\,,
\end{equation}
\begin{equation}\label{g}
g_{\ast}(n)=g_{_{-}}+\big(1-2^{-(2n+3)}\big)g_{_{+}}\,,
\end{equation}
with $\Gamma(y)$ and $\zeta(y)$ being the Gamma and Riemann Zeta
functions respectively.

The associated energy and entropy densities of the system can be
obtained from their definitions
\begin{eqnarray}\label{rho-def}
\rho(T)=T^2\frac{\partial}{\partial T}\bigg(\frac{\mathrm{P}}{T}
\bigg),\hspace{1.5cm}s=\frac{\rho+{\mathrm{P}}}{T}\,,
\end{eqnarray}
with results
\begin{equation}\label{rho}
\rho(T)=\rho_0(T)\left[1-\Big(1+\frac{2n}{3}\Big)\chi(n)\frac{g_{\ast}(n)
}{g_{0\ast}}\beta_0^{2n}\bigg(\frac{T}{T_{\rm Pl}}\bigg)^{2n}\right]\,,
\end{equation}
\begin{equation}\label{s}
s(T)=s_0(T)\left[1-\Big(1+\frac{n}{2}\Big)\chi(n)\frac{g_{\ast}(n)}{
g_{0\ast}}\beta_0^{2n}\bigg(\frac{T}{T_{\rm Pl}}\bigg)^{2n}\right]\,,
\end{equation}
where $\rho_0(T)=3{\mathrm{P}}_0=\frac{\pi^2g_{0\ast}}{30}\,T^4$ and
$s_0(T)=\frac{
\rho_0+{\mathrm{P}}_0}{T}=\frac{2\pi^2g_{0\ast}}{45}\,T^3$ are the
standard energy and entropy densities respectively. The relation
(\ref{rho}) coincides with the result obtained in Ref.
\cite{Maziashvili} when $n=1$. It is important to note that the
GUP-thermodynamical quantities (\ref{P}), (\ref{rho}) and (\ref{s})
are not positive definite, but all of them are positive for
$T<\left[\frac{\pi^4(g_{0\ast}/g_{\ast}(n))}{45
\Gamma(2n+4)\zeta(2n+4)}\right]^{\frac{1}{2n}}\frac{T_{\rm
Pl}}{\beta_0}$. In the following, we will see that the
thermodynamical quantities approach to a maximum value below this
temperature. We show that this corresponds to a nonsingular
radiation dominated universe.
\section{Cosmological implications}
With the use of the deformed pressure (\ref{P}) and energy density
(\ref{rho}), we can easily consider the GUP effects on the evolution
of the early universe. To do this, note that the energy-momentum
tensor of the radiation $T_{\mu\nu}=(-\rho,{\mathrm P} \delta_{ij})$
will be significantly modified at high temperature regime. For the
spatially flat Friedmann-Robertson-Walker universe with metric
$g_{\mu\nu}= (-1,a^2(t)\delta_{ij})$, the Einstein's equations lead
to
\begin{eqnarray}\label{freidmann}
H^2=\frac{8\pi{G}}{3}\rho\,,
\end{eqnarray}
where $H(t)=\frac{\dot{a}}{a}$ is the Hubble parameter, $a(t)$ is
the scale factor and a dot denotes derivative with respect to the
cosmic time $t$. This relation is nothing but the deformed Friedmann
equation in GUP framework since the energy density is now modified
through the relation (\ref{rho}). The conservation of the
energy-momentum tensor $\nabla_{\mu} T^\mu_\nu=0$ also gives
\begin{equation}\label{Cons-rho}
\dot{\rho}+3H(\rho+{\mathrm P})=0\,,
\end{equation}
which is also deformed in GUP setup since the energy density and
pressure are now given by the relations (\ref{rho}) and (\ref{P})
respectively. Furthermore, bearing in the mind the well-known
adiabatic condition
\begin{eqnarray}\label{adiabatic}
S=s\,a^3=\mbox{constant}\,,
\end{eqnarray}
the total entropy, $S$, of the universe is assumed to be constant in
standard model of cosmology. Using the definitions (\ref{rho-def}),
it is not difficult to show that the relations (\ref{Cons-rho}) and
(\ref{adiabatic}) are nothing other than the manifestation of the
first law of thermodynamics. Substituting the GUP-modified energy
density (\ref{rho}) into the relation (\ref{freidmann}) gives the
deformed Friedmann equation in GUP framework as
\begin{eqnarray}\label{freidmann-n}
H^2=\frac{8\pi{G}}{3}\rho_0\left[1-\Big(\frac{\rho_0}{\rho_c(n)}\Big)^{
\frac{n}{2}}\right]\,,
\end{eqnarray}
in which we have defined the critical energy density
\begin{equation}\label{rho-c}
\rho_c(n)=\left[\frac{\pi^4(g_{0\ast}/g_{\ast}(n))}{45\Gamma(2n+4)
\zeta(2n+4)}\right]^{\frac{2}{n}}\,\rho_{_{\rm Pl}}\,,
\end{equation}
with $\rho_{_{\rm Pl}}=\frac{\pi^2g_{0\ast}}{30}\frac{T_{\rm Pl}^4}{
\beta_0^4}$. For the case $n=1$, the GUP-modified Friedmann equation
(\ref{freidmann-n}) will be reduced to the result that is obtained
in Ref. \cite{Ali} in which the authors studied the GUP effects in
the framework of the Hamiltonian formalism of general relativity.
However, it is important to note that while the conservation of
energy remains unchanged in their setup, it will be modified in our
formulation through the relation (\ref{Cons-rho}).

Now, we note that at the temperature
$T_{*}=\left[\frac{2\pi^4(g_{0\ast}/g_{\ast}(n))}{45(2+n)
\Gamma(2n+4)\zeta(2n+4)}\right]^{\frac{1}{2n}}\frac{T_{\rm
Pl}}{\beta_0}$, the energy density (\ref{rho}) reaches to its
maximum value $\frac{\rho_c}{2} $. The entropy density (\ref{s})
also has maximum at this temperature and thus, according to the
adiabatic condition (\ref{adiabatic}) the scale factor will have a
minimum as
\begin{equation}\label{a-min}
a_{\min}=\left(S\left[\frac{45(3+2n)}{4n\pi^2g_{0\ast}}\right]^{\frac{1}{
3}}\left[\frac{45g_{\ast}(n)}{2\pi^4g_{0\ast}}(2n+1)\Gamma(2n+4)\zeta(2n+4)
\right]^{\frac{1}{2n}}\right)\,\beta_0l_{\rm Pl}.
\end{equation}
Therefore, in the case where GUP considerations are taken into
account, we have a nonsingular cosmology in which the minimum of its
scale factor occurs at the temperature $T_{*}$ for which
$\rho_0=\Big(1+\frac{n}{2} \Big)^{-\frac{2}{n}}\rho_c(n)$. This
means that the evolution of the scale factor based on the quantum
gravity effects, shows a bouncing like behavior in such way that the
classical big bang singularity will be replaced by a minimal size of
the scale factor.

\subsection{The case $n=2$}
Let us now take a look at the special case $n=2$ for which the
modified Friedmann equation
\begin{eqnarray}\label{freidmann-2}
H^2=\frac{8\pi{G}}{3}\rho_0\left(1-\frac{\rho_0}{\rho_c(2)}\right)\,,
\end{eqnarray}
is very similar to that arises from loop quantum cosmology
\cite{LQC,RLQC}. However, apart from this similarity, there is a
crucial difference between them due to the holonomy corrections (see
below). Here, the critical energy density $\rho_c(2)$ is defined
from (\ref{rho-c}), that is
\begin{equation}\label{rho-c2}
\rho_c(2)=\left(\frac{g_{0\ast}}{g_{\ast}(2)}\right)\frac{
\rho_{_{\rm Pl}}}{24\pi^4}\,,
\end{equation}
with $g_{\ast}(2)=g_{_{-}}+\frac{127}{128}g_{_{+}}$ from (\ref{g}).
In this case, the GUP-deformed energy density (\ref{rho}) approaches
to its maximum at $\rho_0=\frac{\rho_c(n)}{2}$ corresponds to the
minimum value of the scale factor as
\begin{equation}\label{a-min2}
a_{\min}=S\left(\frac{3\pi\times{7!!}}{4g_{0\ast}}\right)^{\frac{1}{3}}
\left(5!!\times\frac{g_{\ast}(2)}{g_{0\ast}}\right)^{\frac{1}{4}}\,
\beta_0l_{\rm Pl}.
\end{equation}
It is important to note that while the standard conservation
relation $\dot{\rho}_0+3H(\rho_0+{\mathrm P}_0)=0$ keeps its form in
loop quantum cosmology, it is not the case in GUP formalism since
here we have
\begin{equation}\label{Cons-2}
\dot{\rho}_0+3H(\rho_0+{\mathrm P}_0)=2\frac{\rho_0}{\rho_c}
\Big(\dot{\rho}_0+\frac{12}{7}H\rho_0\Big)\,,
\end{equation}
in which we have used the relations (\ref{P}) and (\ref{rho}). This
is because of the fact that, while the geometric part of the
Einstein's equation gets holonomy corrections in loop quantum
cosmology, by the GUP setting the matter part (energy-momentum
tensor) will be modified. The relation (\ref{Cons-2}) also shows
that the matter may interact with itself in GUP framework.

\section{Summary}
Quantum gravity proposal suggests the existence of a minimal
measurable length below which no other length can be observed.
Taking a minimal length scale into account immediately leads to the
modification of the standard uncertainty principle. In this paper,
we first formulated a generalized uncertainty principle which
supports the existence of a minimal length through the nonzero
uncertainty in position measurement. For a statistical system, we
obtained the associated deformed density of states and then
implemented it in order to explore the minimal length effects on the
thermodynamics of the radiation dominated universe. We showed that
the GUP corrections to the pressure, energy density and entropy
density of the universe yield to a maximum value for the energy and
entropy densities at high temperature regime when the temperature
becomes of the order of Planck temperature. By utilizing the
modified thermodynamical quantities for the energy-momentum tensor,
we obtained the modified Friedmann equation which seems to be
similar to the Friedmann equation in loop quantum cosmology. We saw
that the modifications due to the GUP considerations make a bouncing
like behavior for the scale factor free of the classical big bang
singularity. In addition to singularity avoidance, the appearance of
a minimal size for the scale factor in the presented model is also
interesting in its nature due to prediction of a minimal size for
the corresponding universe. It is well-known that the idea of
existence of a minimal length in nature is supported by almost all
candidates of quantum gravity. Finally, we have noticed a difference
between loop quantum cosmology and GUP results. While in the first
theory the geometric part of the Einstein's equations gets holonomy
corrections, the matter part is modified in the second. Thus,
although the energy conservation relation remains unchanged in loop
quantum cosmology, we showed it will be modified in GUP setup.


\begin{thebibliography}{}

\bibitem{Inflation} A. H. Guth, Phys. Rev. D
{\bf 23} (1981) 347\\A. D. Linde, Phys. Lett.
B {\bf 108} (1982) 389\\A. Albrecht and P. J.
Steinhardt, Phys. Rev. Lett. {\bf 48} (1982)
1220.

\bibitem{Ashtekar} A. Ashtekar, Phys. Rev. Lett.
{\bf 57} (1986) 2244\\A. Ashtekar, Phys. Rev. D
{\bf 36} (1987) 1587.

\bibitem{LQG} C. Rovelli and L. Smolin, Nucl.
Phys. B {\bf 331} (1990) 80\\A. Ashtekar and
J. Lewandowski, Class. Quantum Grav. {\bf 14}
(1997) A55\\A. Ashtekar and J. Lewandowski,
Adv. Theor. Math. Phys. {\bf 1} (1998) 388\\A.
Ashtekar and J. Lewandowski, Class. Quantum
Grav. {\bf 21} (2004) R53.

\bibitem{RLQG} C. Rovelli, Living Rev. Rel.
{\bf 1} (1998) 1\\C. Rovelli, {\it Quantum
Gravity}, Cambridge University Press, Cambridge,
UK, (2004)\\T. Thiemann, {\it Modern Canonical
Quantum General Relativity}, Cambridge
University Press, Cambridge, UK, (2007).

\bibitem{LQC} M. Bojowald, Phys. Rev. Lett.
{\bf 86} (2001) 5227\\M. Bojowald, Class.
Quantum Grav. {\bf 19} (2002) 2717\\A.
Ashtekar, T. Pawlowski and P. Singh, Phys.
Rev. Lett. {\bf 96} (2006) 141301\\A.
Ashtekar, T. Pawlowski and P. Singh, Phys. Rev.
D {\bf 74} (2006) 084003\\A. Ashtekar, T.
Pawlowski and P. Singh, Phys. Rev. D {\bf 73}
(2006) 124038.

\bibitem{RLQC} A. Ashtekar, M. Bojowald and J.
Lewandowski, Adv. Theor. Math. Phys. {\bf 7}
(2003) 233\\M. Bojowald, Living Rev. Rel.
{\bf 11} (2008) 4\\A. Ashtekar and P. Singh,
Class. Quantum Grav. {\bf 28} (2011) 213001.

\bibitem{LQG-VO} C. Rovelli and L. Smolin,
Nucl. Phys. B {\bf 442} (1995) 593\\R. Loll,
Phys. Rev. Lett. {\bf 75} (1995) 3048\\R.
Loll, Nucl. Phys. B {\bf 460} (1996) 143\\
J. Brunnemann and T. Thiemann, Class.
Quantum Grav. {\bf 23} (2006) 1289.

\bibitem{String} D. J. Gross and P. F. Mende,
Nucl. Phys. B {\bf 303} (1988) 407\\D. Amati,
M. Ciafaloni and G. Veneziano, Phys. Lett. B
{\bf 216} (1989) 41\\K. Konishi, G. Paffuti
and P. Provero, Phys. Lett. B {\bf 234}
(1990) 276\\L. Garay, Int. J. Mod. Phys. A
{\bf 10} (1995) 145.

\bibitem{GUP} M. Maggiore, Phys. Lett. B
{\bf 304} (1993) 65\\M. Maggiore, Phys. Lett.
B {\bf 319} (1993) 83\\M. Maggiore, Phys.
Rev. D {\bf 49} (1994) 5182.

\bibitem{GUPUV} A. Kempf, G. Mangano and R.
B. Mann, Phys. Rev. D {\bf 52} (1995) 1108.

\bibitem{GUPUV2} A. Kempf and G. Mangano,
Phys. Rev. D {\bf 55} (1997) 7909\\K.
Nozari and A. Etemadi, Phys. Rev. D
{\bf 85} (2012) 104029.

\bibitem{GUP-THR} S. K. Rama, Phys. Lett.
B {\bf 519} (2001) 103\\M. Lubo, Phys. Rev.
D {\bf 68} (2003) 125004\\K. Nozari and B.
Fazlpour, Gen. Reltiv. Grav. {\bf 38}
(2006) 1661\\P. Wang, H. Yang and X. Zhang,
JHEP {\bf 08} (2010) 043\\P. Wang, H. Yang
and X. Zhang, Phys. Lett. B {\bf 718}
(2012) 265.

\bibitem{GUP-DOS} L. N. Chang, D. Minic,
N. Okamura and T. Takeuchi, Phys. Rev. D
{\bf 65} (2002) 125028\\A. F. Ali, Class.
Quantum Grav. {\bf 28} (2011) 065013\\P.
Pedram, Phys. Lett. B {\bf 718} (2012)
638.

\bibitem{GUP-Fityo} T. Fityo, Phys. Lett.
A {\bf 372} (2008) 5872.

\bibitem{Maziashvili} D. Mania and M.
Maziashvili, Phys. Lett. B {\bf 705}
(2011) 521.

\bibitem{GUP-THR2} B. Vakili and M. A.
Gorji, J. Stat. Mech. (2012) P10013.

\bibitem{GUP-NS} M. V. Battisti and G.
Montani, Phys. Lett. B {\bf 656} (2007)
96.

\bibitem{Ali} A. F. Ali and B. Majumder,
Class. Quantum Grav. {\bf 31} (2014)
215007

\bibitem{Ali2} A. Awad and A. F. Ali,
JHEP {\bf 1406} (2014) 093.

\bibitem{experiment} S. Das and E. C.
Vagenas, Phys. Rev. Lett. {\bf 101}
(2008) 221301\\ P. Pedram, K. Nozari
and S. H. Taheri, JHEP {\bf 1103}
(2011) 093\\S. Jalalzadeh, M. A. Gorji,
K. Nozari, Gen. Reltiv. Grav. {\bf 46}
(2014) 1632.

\bibitem{DOS-LT} M. A. Gorji, K. Nozari
and B. Vakili, Phys. Rev. D {\bf 89}
(2014) 084072.


\end{thebibliography}
\end{document}